\documentclass[reprint, amsmath,amssymb,aps, onecolumn]{revtex4-2}

\usepackage{graphicx}
\usepackage{subfigure}
\usepackage{subfloat}
\usepackage{amssymb}
\usepackage{amsmath}
\usepackage{multirow}
\usepackage{dcolumn}
\usepackage{bm}
\usepackage{wasysym}
\usepackage{cancel}

\usepackage{txfonts}
\usepackage{lineno}

\begin{document}

\title{Magnetic Effect on Potential Barrier for Nucleosynthesis}
\thanks{Magnetic Effect on Potential Barrier for Nucleosynthesis}%

\author{Kiwan Park}
 \affiliation{Department of Physics, OMEG, Soongsil University, 369, Sangdo-ro, Dongjak-gu, Seoul 06978 Republic of Korea\\ pkiwan@ssu.ac.kr\\}
\author{Yudong Luo}
\affiliation{School of physics, Kavli Institute for Astronomy and Astrophysics, Peking University, Beijing 100871, China\\}

\author{Toshitaka Kajino}
\affiliation{School of Physics, International Research Center for Big-Bang Cosmology and Element Genesis, and Peng Huanwu Collaborative Center for Research and Education, Beihang University, Beijing 100083, China}
\affiliation{Division of Science, National Astronomical Observatory of Japan, 2-21-1 Osawa, Mitaka, Tokyo 181-8588, Japan}
\affiliation{Graduate School of Science, The University of Tokyo, 7-3-1 Hongo, Bunkyo-ku, Tokyo 113-033, Japan\\}

\date{\today}

\begin{abstract}
We investigate the impact of magnetic fields on the potential barrier between two interacting nuclei. We addressed this by solving the Boltzmann equation and Maxwell's theory in the presence of a magnetic field, resulting in the determination of magnetized permittivity. Additionally, we derived the magnetized Debye potential, which combines the conventional Debye potential with an additional magnetic component. We then compared the Boltzmann approach with the Debye method. Both methods consistently demonstrate that magnetic fields increase permittivity. This enhanced permittivity leads to a reduction in the potential barrier, consequently increasing the reaction rate for nucleosynthesis. Furthermore, the dependence on temperature and electron density in each approach is consistent. Our findings suggest that magnetized  plasmas from the hot Big-Bang Universe to the Solar interior have played a crucial role in nucleosynthesis.
\end{abstract}

\maketitle

\section{Introduction}
\begin{figure*}
    {
   \subfigure[$\epsilon(k,B)$]{
     \includegraphics[width=9.2 cm]{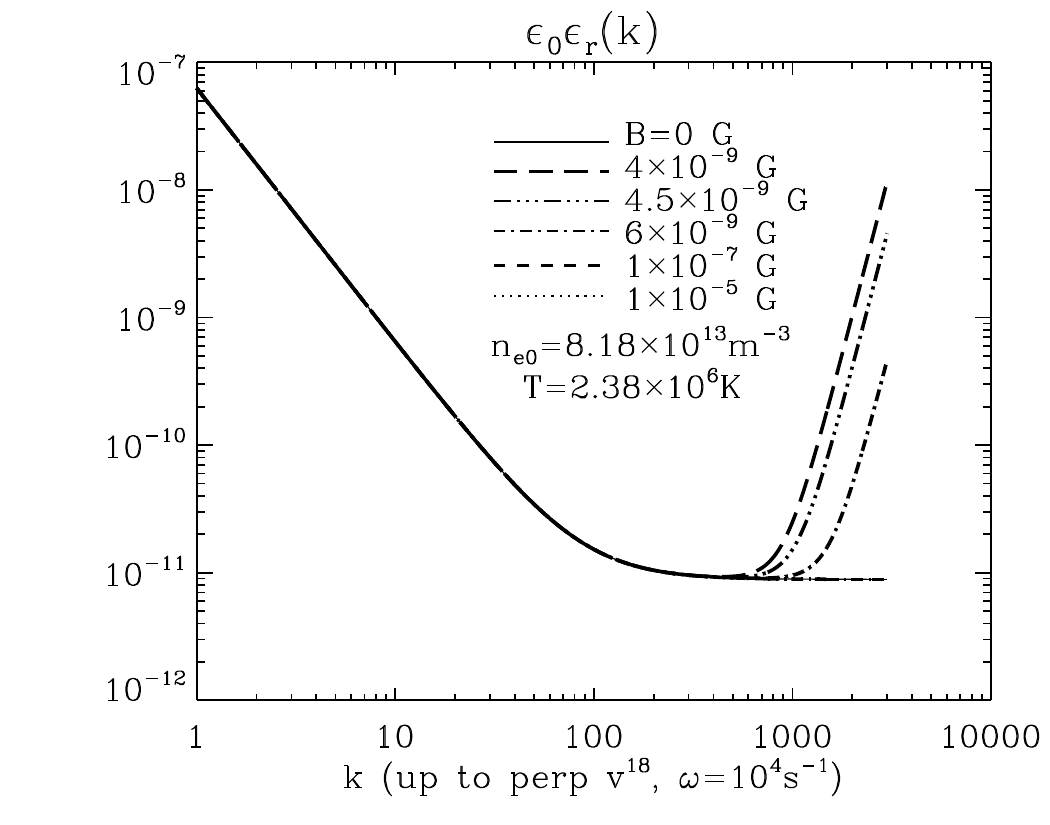}
     \label{f1}
    }\hspace{-13 mm}
   \subfigure[ $\epsilon(k, T)$ ]{
   \includegraphics[width=9.2 cm]{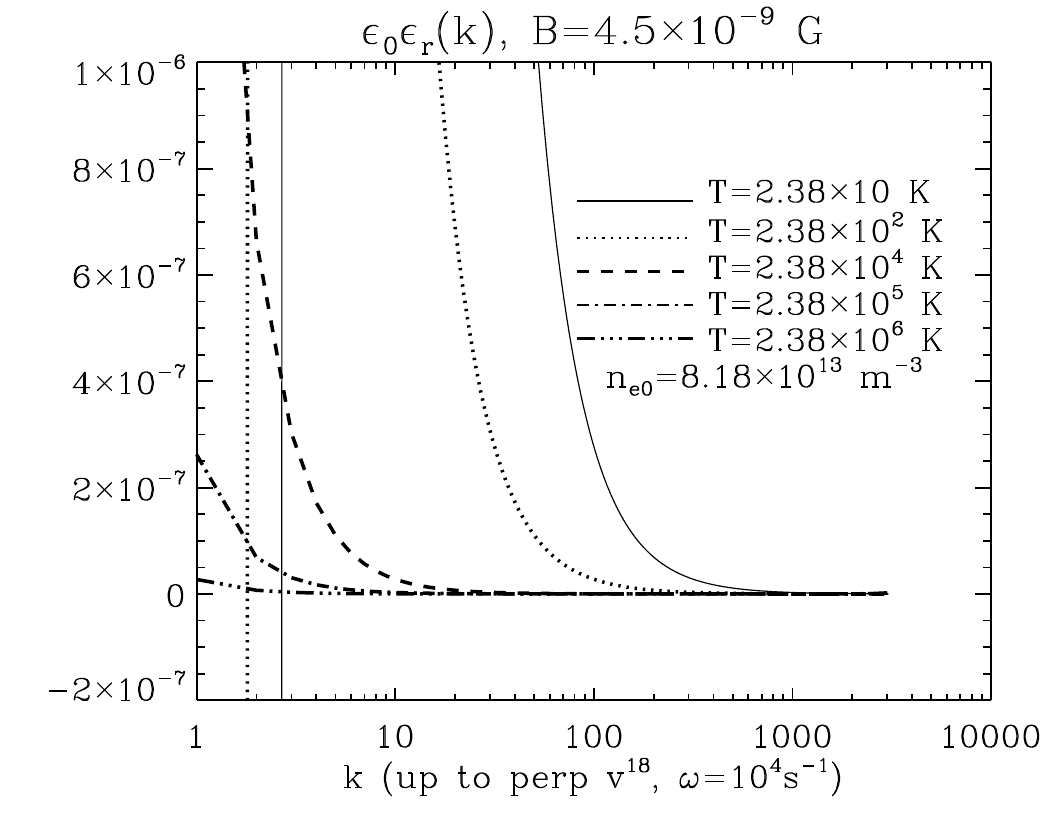}
     \label{f21}
   }\hspace{-13 mm}
   \subfigure[$\epsilon(k,n_{e0})$]{
   \includegraphics[width=9.2 cm]{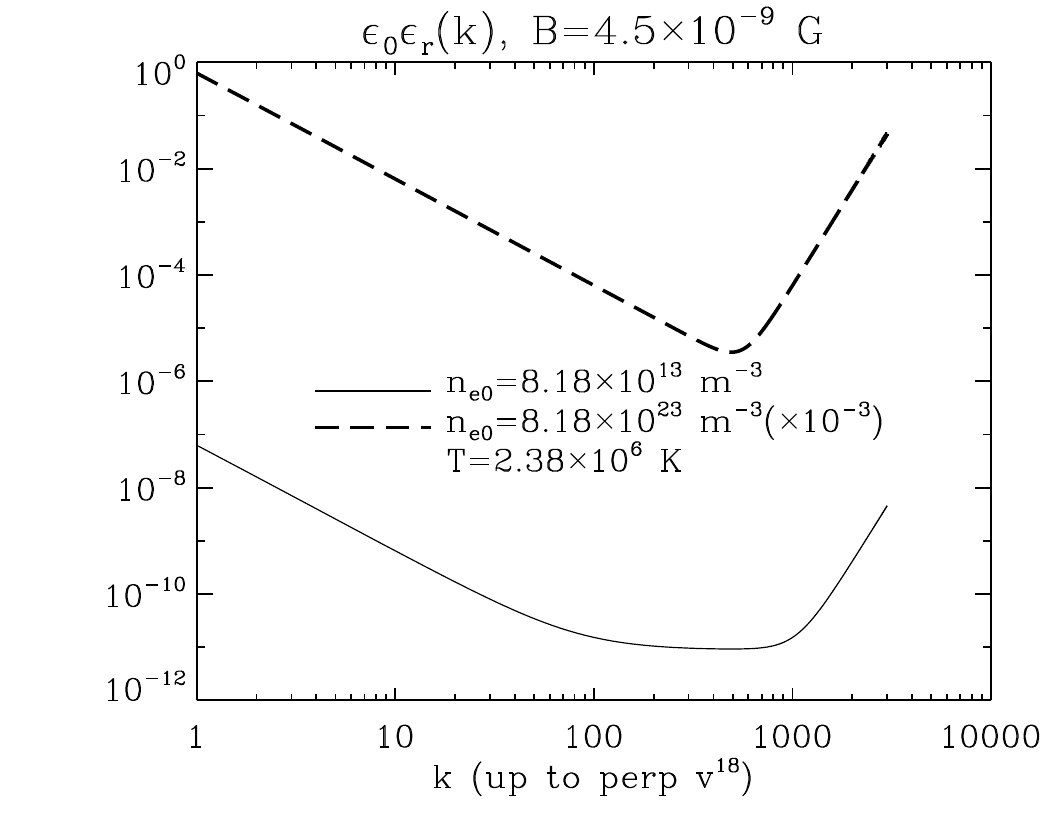}
     \label{f22}
   }\hspace{-13 mm}
   \subfigure[ $\epsilon(r/\lambda_D,B)$ ]{
     \includegraphics[width=9.2 cm]{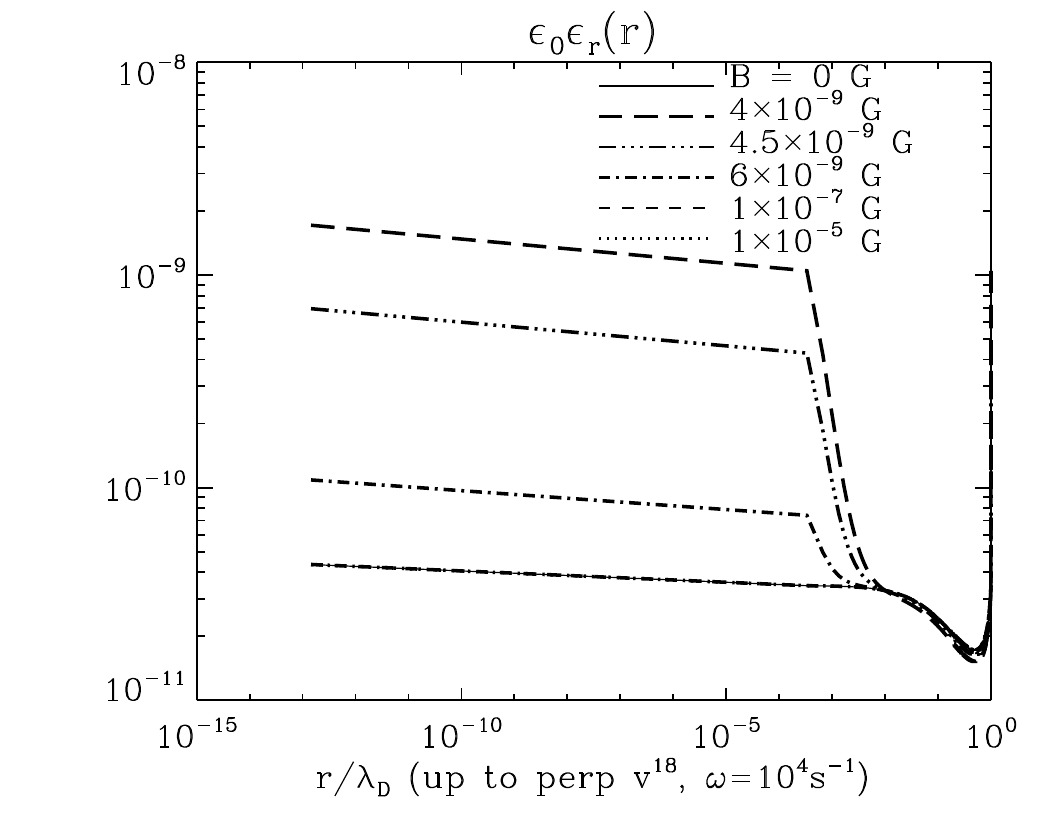}
     \label{f2}
   }
}
\caption{(a) $\epsilon(k)$ in Fourier space (b) $\epsilon(k)$ depending on T (inversely proportional) (c) $\epsilon(k)$ depending on $n_{e0}$ (proportional) (d) $\epsilon(r/\lambda_D)$ in real space. {The magnetic fields stronger than $>4\times 10^{-9}G$ render the second and third term $\sim (m_e/B)^{n+2s}$ in Eq.~(\ref{final_epsilon_numerical}) relatively negligible. Physically, the superposition of electrons, resulting from their cyclotron rotation around the magnetic field ($r\sim 1/B$), is inversely proportional to the strength of magnetic field. The increasing superposing effect due to the weaker magnetic field more efficiently shields the electric field from the nucleus. This is why the application of a increasing magnetic field converges to the non-magnetized case.}}
\end{figure*}

\begin{figure*}
    {
   \subfigure[$\frac{e}{4\pi \epsilon r}$]{
     \includegraphics[width=9.2 cm]{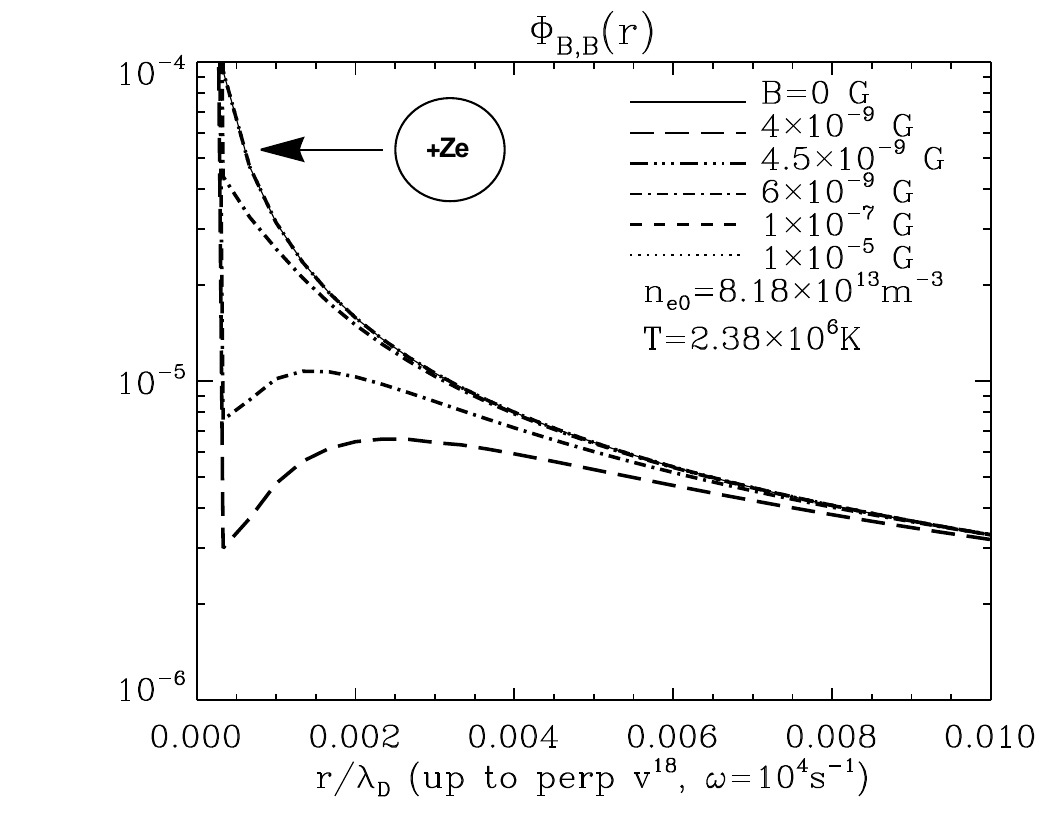}
     \label{f3}
   }\hspace{-13 mm}
   \subfigure[$exp\bigg(-\frac{Z_1Z_2e^2}{2\epsilon \hbar}\sqrt{\frac{\mu}{2E}}\bigg)$, $Z_1=Z_2=1$ ]{
     \includegraphics[width=9.2 cm]{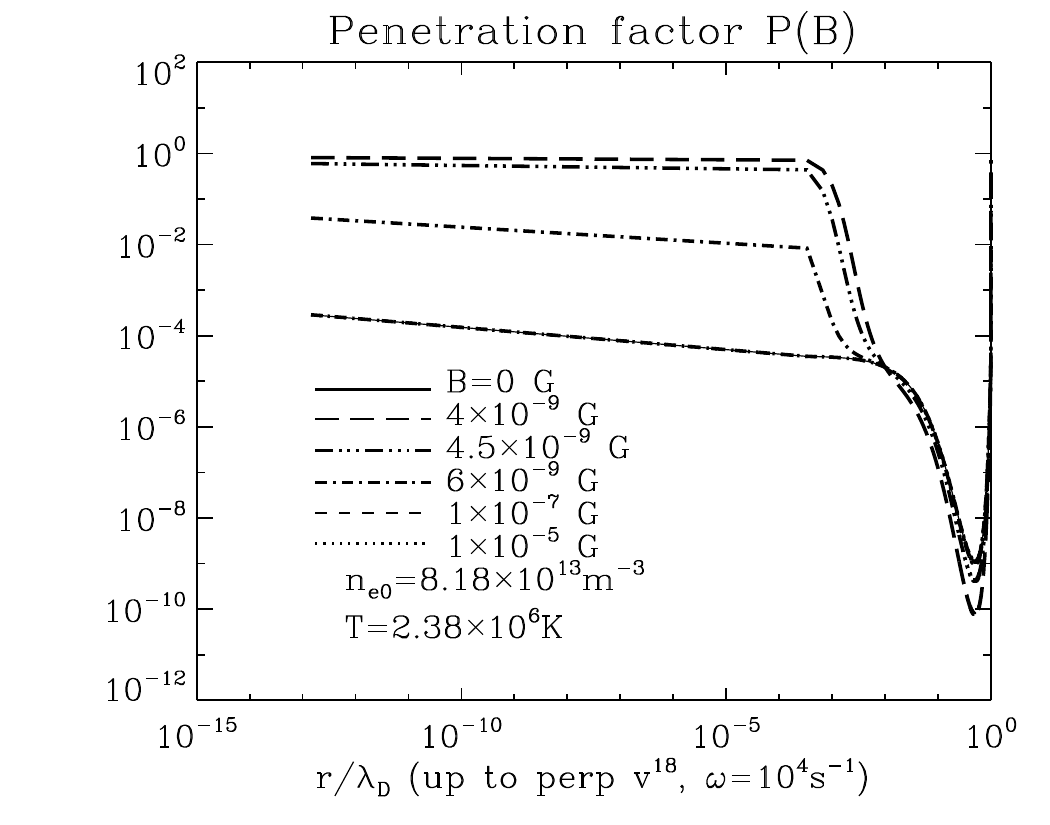}
     \label{f4}
   }
}
\caption{(a) Potential barrier with $\epsilon(r/\lambda_D)$ with the magnetic field (b) Penetration factor with $\epsilon(r/\lambda_D)$ with magnetic field}
\end{figure*}

\begin{figure*}
{
   \subfigure[]{
     \includegraphics[width=9.2 cm]{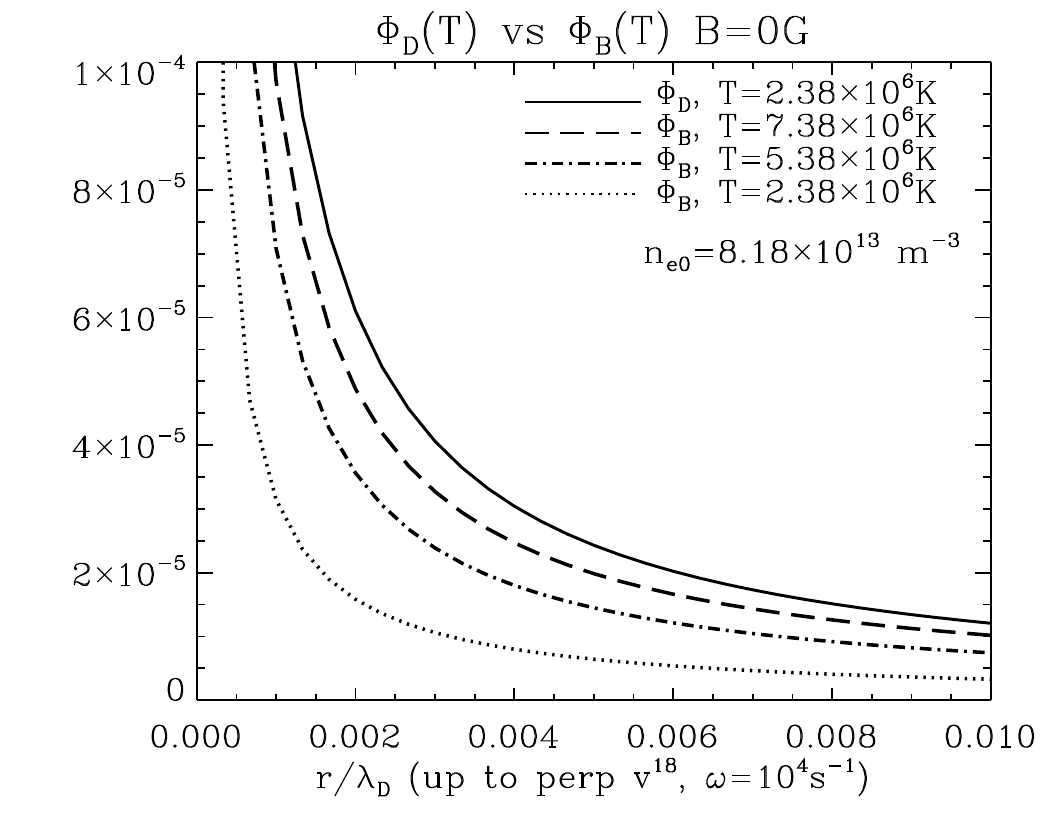}
     \label{f41}
}\hspace{-13 mm}
   \subfigure[]{
     \includegraphics[width=9.2 cm]{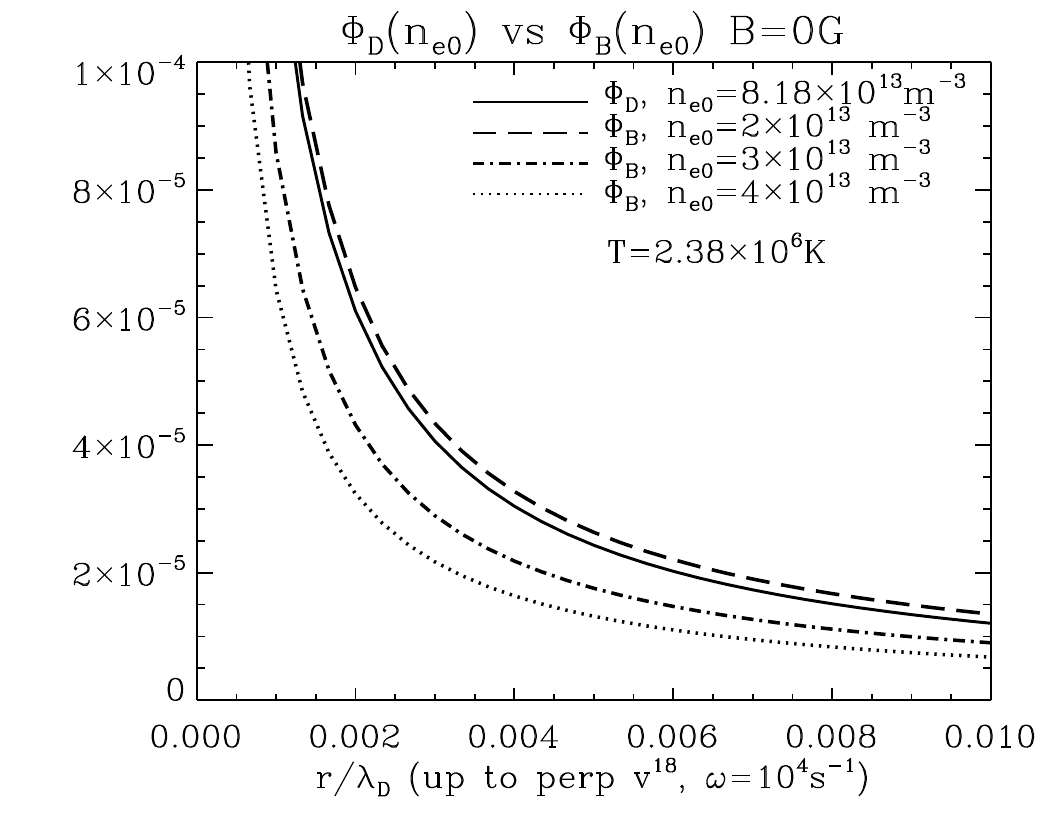}
     \label{f42}
   }
   \subfigure[]{
     \includegraphics[width=9.2 cm]{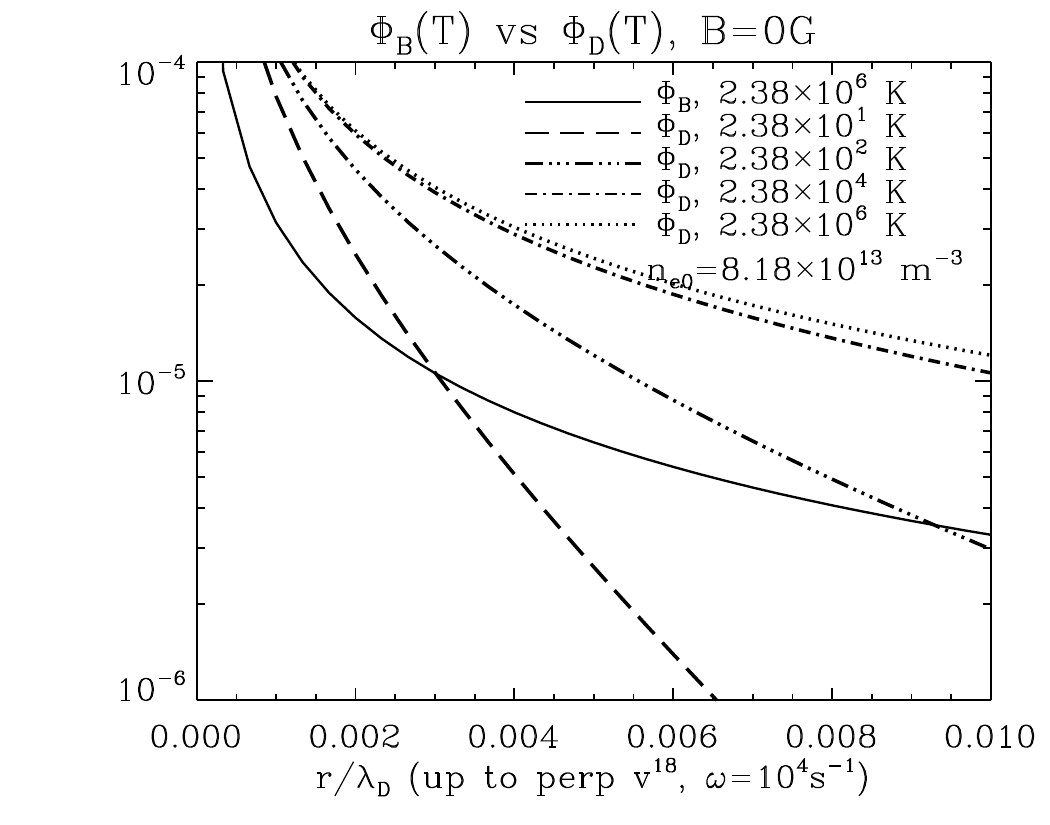}
     \label{f5}
}\hspace{-13 mm}
   \subfigure[]{
     \includegraphics[width=9.2 cm]{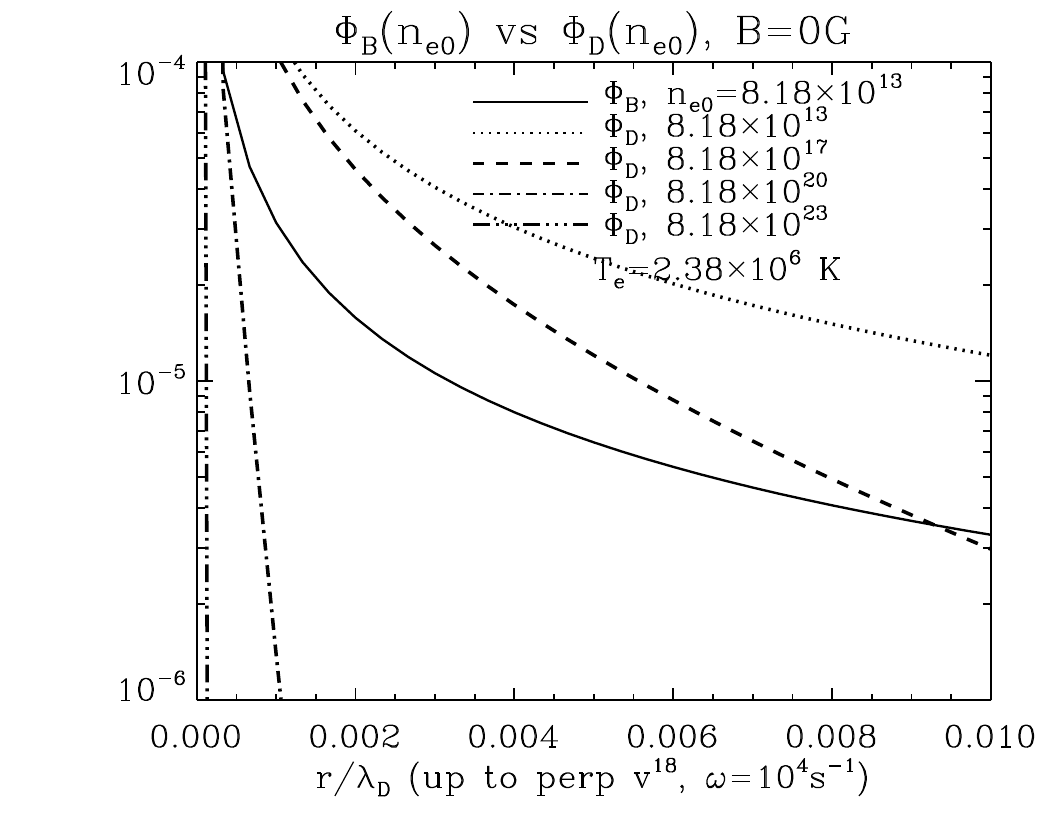}
     \label{f6}
   }
}
\caption{{(a), (b) The solid lines indicate potential energy $\Phi_D$ without magnetic field. Other lines indicate Boltzmann potential energy $\Phi_B$ depending on $T$ and $n_{e0}$. In (c), (d), the solid lines in indicate potential energy $\Phi_B$ without magnetic field. Other lines indicate unmagnetized Debye potential energy $\Phi_D$ with various $T$ and $n_{e0}$. Both $\Phi_B$ and $\Phi_D$ grow in proportion to the temperature, but they are inversely proportional to $n_{e0}$. {The validity of Debye potential is guaranteed with $\Phi\ll k_BT/e\sim2\times 10^{-3}-2\times 10^2$.}}}
\end{figure*}

\begin{figure*}
{
   \subfigure[]{
     \includegraphics[width=9.2 cm]{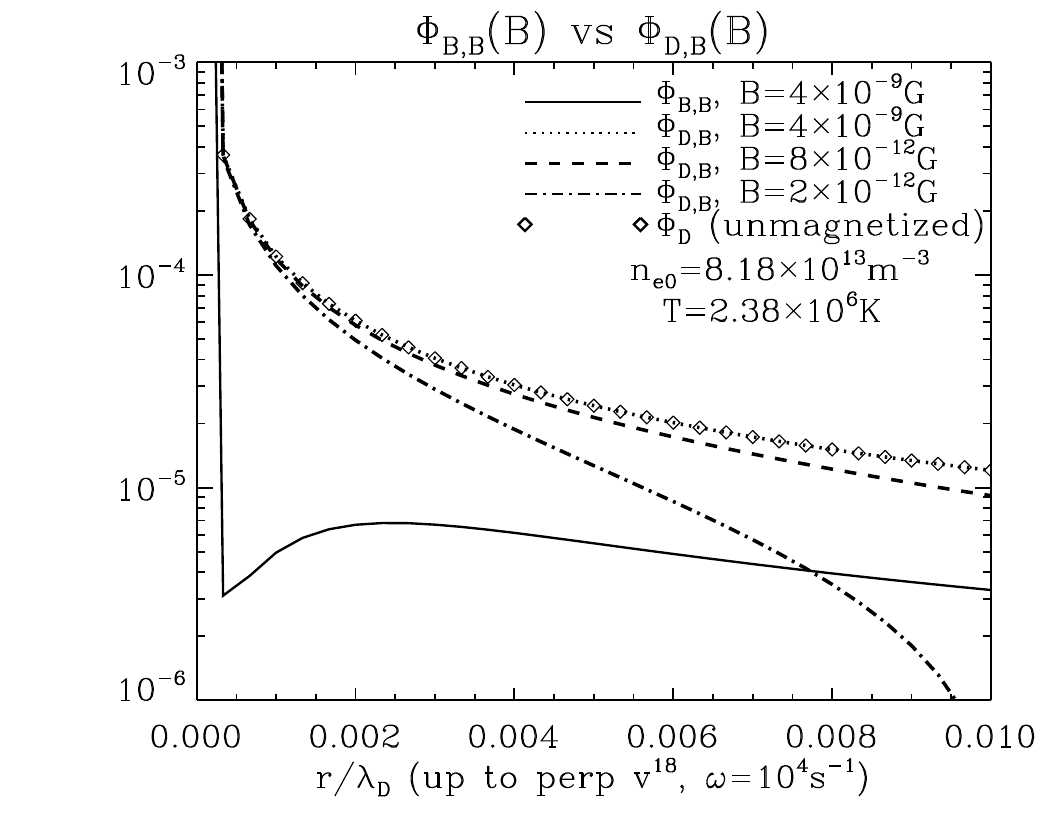}
     \label{f7}
}\hspace{-13 mm}
   \subfigure[]{
     \includegraphics[width=9.2 cm]{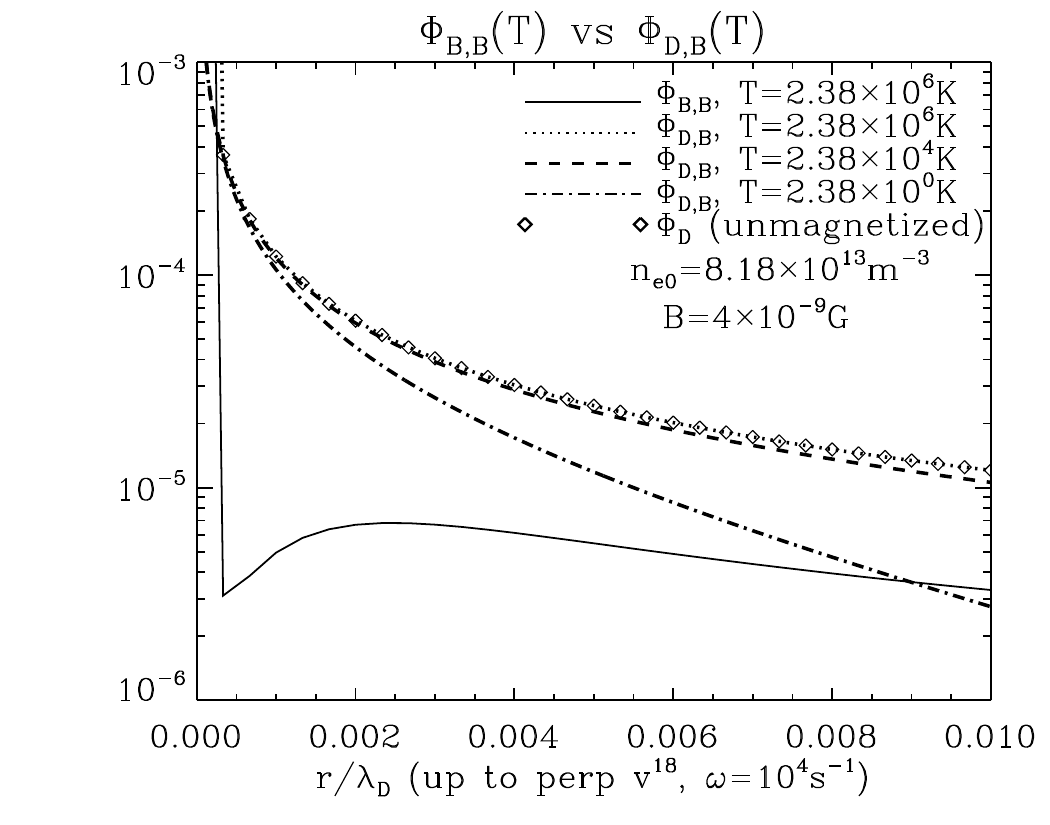}
     \label{f8}
   }
}
\caption{The solid lines in (a) and (b) represent the profile of potential energy $\Phi_{\text{B,B}}$ obtained from the magnetized Boltzmann equation, Eq. (\ref{final_epsilon_numerical}), with $B=4\times 10^{-9}\text{G}$, $T=2.38\times 10^{6}\text{K}$, and $n_{0e}=8.18\times 10^{13}\text{m}^{-3}$. In (a), the magnetized Debye potential $\Phi_{D,B}(B)$ from Eq. (\ref{derivation_of_new_Debye_potential_12}), with varying magnetic field and fixed $n_{e0}$ and $T$, is compared. In (b), $\Phi_{D,B}(T)$, with varying temperature and fixed $n_{e0}$ and fixed magnetic field, is compared. The dependence on $n_{e0}$ is not considered in this comparison. {And, the diamond lines represent unmagnetized Debye potential energy $\Phi_D$ with the same $T$ and $n_{e0}$ as $\Phi_{D,B}$ of the dotted line. This shows that the strong magnetic effect on the electron sphere is negligible.}}
\end{figure*}

Magnetic fields ($B$) and plasmas are prevalent throughout the Universe. However, despite extensive research, the role of magnetic fields in the evolution of celestial plasma systems remains only partially understood. On a macroscopic scale, magnetic fields are associated with the instability of plasma systems. Additionally, they transport angular momentum, resulting in a deceleration of rapid collapses and facilitating continued accretion. Nevertheless, our comprehension of these phenomena only skims the surface of the magnetic field's role. In reality, our knowledge of the microscopic effects of magnetic fields, especially their impact on the synthesis of fundamental elements in the Universe, is quite limited.\\

Nucleosynthesis remains a partially understood phenomenon, and there have been longstanding debates about the impact of (non-magnetized) plasma on nuclear fusion. Furthermore, research on the magnetic field's influence on nucleosynthesis has been quite limited. We believe that the main reasons for overlooking the magnetic field are the complexity it introduces and its exceedingly weak strength in the early Universe. Nevertheless, a model that considers a universally magnetized plasma environment is crucial for achieving a comprehensive understanding. In this context, our focus is to investigate the effects of magnetic fields on nuclear reactions.\\


Nucleosynthesis proceeds through a series of processes including the proton-proton (pp) chain, CNO cycle, triple-alpha reaction, and so on. The reaction rate is represented as
\begin{eqnarray}
R\sim \langle \sigma v \rangle &=& \frac{2^{3/2}}{\sqrt{\pi \mu}}\frac{1}{(k_BT)^{3/2}}\int^{\infty}_0 S(E)
\,exp\bigg[-\frac{E}{k_BT}-\frac{Z_1Z_2e^2}{2\epsilon\hbar}\bigg(\frac{\mu}{2 E}\bigg)^{1/2}\bigg]dE \label{Reaction_suppression1}\\
&\sim& \frac{S(E_0)}{T^{2/3}}exp\bigg[  -\bigg(\frac{\mu Z_1^2 Z_2^2}{m_p}\, \frac{7.726\times 10^{10} K}{\epsilon_r T}\bigg)^{1/3} \bigg].
\label{Reaction_suppression2}
\end{eqnarray}
Here $\epsilon$, $\mu$, $m_p$, $E$, and $E_0$ are respectively `permittivity', `reduced mass', `proton mass', and `thermal energy $E=3k_BT/2$', and Gamow peak energy at temperature $T$. And $\epsilon_r(=\epsilon/\epsilon_0)$ is the relative permittivity, where $\epsilon_0(=8.85\times 10^{-12} F/m)$ is the permittivity in free space. We hereafter call the exponential function of Eq.(\ref{Reaction_suppression2}) the 'penetration factor' $P(E)$ at the Gamow peak energy at $E=E_0$. As this formula shows, the nucleosynthesis requires a significant amount of energy to overcome the Coulomb barrier between two fusing ions. For instance, in the Solar core ($T\sim 10^7 K$) and the early Universe after the Big Bang ($t\sim1-10^2s,\,T\sim 10^{10} K$), the reaction rates for synthesizing deuterium $\mathrm{^2H}$ in the initial step of the proton-proton (pp) chain are suppressed by $1.53\times 10^{-7}$ and $0.21$, respectively. Furthermore, in the subsequent step involving $\mathrm{^3He}$, these rates are further reduced by $1.24\times 10^{-12}$ and $0.064$.\\

{The thermodynamic conditions of the plasma(i.e., density, pressure, temperature) are the key parameters for both astrophysical evolution and the nuclear reactions for nucleosynthesis studies.} In particular, the enhanced electron density decreases the potential barrier, thereby increasing the penetration factor and reaction rate. However, modifying the electron density in the plasma is a somewhat artificial laboratory process that deviates from the natural state. Therefore, several models that enhance the reaction rate without artificial manipulations have been proposed. One such model is the screening effect, which is attributed to the presence of densely packed electrons surrounding the ions. The screening effect can reduce the Coulomb barrier and enhance the reaction rate (refer to Debye-H$\mathrm{\ddot{u}}$ckel screening in \cite{2003phpl.book.....B}).\\

Salpeter \cite{1954AuJPh...7..373S} proposed the concept of electron screening surrounding fusing nuclei, which includes the static interaction among the fusing ions and electron clouds. Subsequently, several studies and suggestions were based on this ground breaking work. Bahcall et al. \cite{1998PhRvC..57.2756B} solved the Debye potential using the WKB approximation, where the Coulomb wave function naturally emerges from Salpeter's formulation\citep{1954AuJPh...7..373S, salpeter69}. And, Gruzinov et al. \cite{1998ApJ...504..996G} calculated the partial differential equation for the electron density matrix in the vicinity of two nuclei. Also, Dewitt et al. \cite{1973ApJ...181..439D} and Br{\"u}ggen et al. \cite{1997ApJ...488..867B} derived the reaction rate based on the free energy between two ions under the assumption of weak screening.   {Brown et al. \cite{1997RvMP...69..411B} developed a comprehensive model for fusion reactions in plasmas. These models  encompass Salpeter's result under suitable conditions.\\}

{Concurrently, however, Salpeter's static screening effect was pointed out to be inappropriate for the dynamic stellar core. Shaviv et al.\cite{1997ASSL..214...43S}, Carraro et al.\cite{1988ApJ...331..565C}, and Hwang et al.\cite{2021JCAP...11..017H} considered the dynamic effect with the different velocities of nuclei and electrons. Opher et al.\cite{2000ApJ...535..473O} statistically reinterpreted Gibbs distribution of particles in plasmas. And, Shaviv et al.\cite{1996ApJ...468..433S} studied the interaction effect of electrons around the fusing nuclei. As these examples show, authors come to their own plasma models to the nucleus reactions with their own backgrounds and models (\cite{2002A&A...383..291B}, and reference therein). However, the effect of background magnetic field on permittivity $\epsilon_r=\epsilon/\epsilon_0$ in the penetration factor $P\sim exp[-g(\epsilon,\,E,\,Z_1,\,Z_2)]$ has not yet been sufficiently studied in more detail (see Eq.(\ref{Reaction_suppression1})).}\\


{Alternatively, at lower temperature case (i.e., non-relativistic), the linear response theory is useful to calculate the dielectric coefficient of the gas \citep{10013390197} which directly determining the weak screening potential. The dominated state is the ground state of the electron gas in the system. For astrophysical the degenerate relativistic electron-positron plasma, \cite{1962NCim...25..428J} calculated the screening potential of a relativistic ultra-degenerate plasma, the later studies indicate that such screening potential could affect $\beta-$decay rate \citep{morita1973beta, Matese:1965zz} as well as the electron capture rate \citep{Glauber:1956zza, Takahashi1978, itoh02, 2010NuPhA.848..454J}.}\\

On the other hand, we briefly mentioned that the impact of the magnetic field on nucleosynthesis has not been thoroughly studied. In fact, the magnetic field is as widespread and ancient as plasma particles, including their (nucleo)synthesis, in the Universe. The magnetic field has existed ubiquitously since the Big Bang until today in various phases of cosmic evolutionary history. In the very early Universe, various quantum fluctuations, such as QCD or phase transitions followed by plasma fluctuation (Biermann battery effect), induced magnetic fields \citep{1950ZNatA...5...65B, 1994PhRvD..50.2421C, 2012ApJ...759...54T}. These primordial magnetic fields (PMF) are inferred to have been very weak ($10^{-62}-10^{-19}G$) compared to the currently observed mean magnetic field strength of order ($10^{-9}-10^{-5}G$)\citep{2012PhR...517..141Y}, which implies various dynamo processes \citep{2005PhR...417....1B, 2012MNRAS.419..913P, 2012MNRAS.423.2120P}. However, even such a weak magnetic field can magnetize the light electrons surrounding the nuclei. Moreover, this weak field, which loosely constrains the charged particles (electrons), has a more efficient effect on perturbing the distribution of electrons compared to a strong magnetic field, {i.e., cyclotron radius $r\sim 1/B$.} Typically, such weak magnetic fields are overlooked in conventional plasma physics. Nonetheless, during the early Universe, where these faint magnetic fields emerged due to quantum fluctuations and plasma fluctuations and where various nuclei were synthesized, the weak fields may have played a pivotal role in enhancing nuclear reactions.\\

{The magnetic fields can alter the electron density that shields the static electric field around heavy nuclei. We emphasize that electrons superposed by the fields exhibit a similar electric field shielding effect as dense electrons. These series of processes reduce the potential barrier between two reacting nuclei. We will demonstrate the impact of the magnetic field on the permittivity, denoted as $\epsilon_r=\epsilon/\epsilon_0$, which, in turn, contributes to the penetration factor $P(E)$ for nucleosynthesis. However, a strong magnetic field suppresses the perturbation of electrons due to its strong constraining effect.\\}

Following this chapter, in chapter 2 we show how to get the permittivity with Boltzmann equation and electromagnetic theory, analytically and numerically. But, here, we will exclude very detailed standard plasma standard theory. In section 3, we show our numerical results for the magnetized permittivity, potential barrier, and penetration factor. In section 4, we derive the magnetized Debye potential in comparison with the potential energy from Boltzmann equation. In section 5, we summarize our work.\\

\begin{table}

\centering

\begin{tabular}{|l|c|c|c|}
  \hline
  Harmonic Oscillator (Bound charge, Eq.(\ref{epsilon_harmonic_oscillator2})) & $\Phi\sim \mathcal{O}(1/n_e)$ & $-$ & $-$ \\ \hline
  Momentum + Current (Eq.(\ref{effective_permittivity})) & $\sim \mathcal{O}(n_e)$ & $-$ & $-$ \\ \hline
  Boltzmann + Current (Bergmann, longitudinal, $v_{th}=0$)  & $\sim \mathcal{O}(n_e)$ & $-$ & $\sim \mathcal{O}(1/B)$ \\ \hline
        (Bergmann, transverse, $\omega=0, v_{th}=0$)        & $\sim \mathcal{O}(1/n_e)$ & $-$ & $\sim \mathcal{O}(B)$ \\ \hline
  Boltzmann + Bound charge (Eq.(\ref{permittivity7})) & $\sim \mathcal{O}(1/n_e)$ & $\sim \mathcal{O}(T)$ & $\sim \mathcal{O}(B)$  \\ \hline
  Magnetized Debye (Eq.(\ref{derivation_of_new_Debye_potential_12})) & $\sim \mathcal{O}(1/n_e)$ & $\sim \mathcal{O}(T)$ & $\sim \mathcal{O}(B)$ \\
  \hline
   \multicolumn{4}{|l|}{
   $R$: reaction rate, $P$: penetration factor, $S(E)$ : slowly varying function, $E$: thermal energy, $E_0$: Gamow peak energy (see Eq.~1, 2 and Fig.~2b),
   } \\

\multicolumn{4}{|l|}{
$\bf E$: electric field (others), $\bf P$: dipole moment (Eq.7), $f(=f_0+f_1)$: distribution function, $f_0,\,f_1$: mean and perturbed distribution function,
   } \\

   \multicolumn{4}{|l|}{
$F_0 (= f_0/n_{e0})$: Maxwell Boltzmann distribution function, $F(r)$: function for the modified potential (Eq~39),
$n_{e0}$: electron density, $n$: index (Eq.~24),
   } \\

   \multicolumn{4}{|l|}{
$\epsilon(=\epsilon_0\epsilon_r)$: permittivity, $\epsilon_0$: permittivity in vacuum, $\epsilon_r$: relative permittivity, $\epsilon_{l}$: relative permittivity parallel to a B field (longitudinal direction),
   } \\

   \multicolumn{4}{|l|}{
$v_{\|},\, v_{\bot}$: velocity parallel and perpendicular to a B field, $\nu_m$: collision frequency, $\bf U$ : mean velocity (Eq.~29), $F(r)$: modified potential factor (Eq.~39),
   } \\

   \multicolumn{4}{|l|}{
$\Phi_B$: Boltzmann potential without $B$, $\Phi_D$: Debye potential without $B$, $\Phi_{B,B}$: magnetized Boltzmann potential, $\Phi_{D,B}$: magnetized Debye potential
   } \\

   \multicolumn{4}{|l|}{
$\alpha=$ {\Large$\frac{k_{\|} v_{\|} - \omega }{\omega_{ce}}$},
    $\beta=$ {\Large$\frac{k_{\bot} v_{\bot}}{\omega_{ce}}$},
    $\omega_{ce}$= {\Large$\frac{eB}{m_e}$} (cyclotron frequency),
    $\omega^2_{pe}$= {\Large$\frac{n_{e0}e^2}  {\epsilon_0m_e}$} (plasma frequency),
    $\lambda^2_D$: {\Large $\frac{\epsilon_0 k_B T_e}{n_e e^2}$} (Debye length)
   } \\

  \hline
  \end{tabular}
\caption{ {$\Phi\sim \mathcal{O}(X)$ or $\mathcal{O}(1/X)$ simply indicates that $\Phi$ grows along with $X$ or $1/X$ nonlinearly.}}
\end{table}

\section{Theoretical Analysis I}
\subsection{Kinetic Approach}
{Statistically, the many replicas of closed structures composed of nuclei and electrons can be regarded as a canonical ensemble system dominated by Hamiltonian dynamics with generalized coordinates `$q_s$' and momentum `$p_s(=m_sv_s+q_sA,\, \mathbf{B}=\nabla \times \mathbf{A})$'. Liouville's theorem indicates that the total time (material) derivative of the density or distribution function in phase space is $zero$ as we move along the trajectory dominated by Hamiltonian dynamics. Therefore, the alteration of the (external) magnetic field on the system results in changes in the distributions of components in an inversely proportional manner.\\

We consider a box $dqdp$ in the phase space with the distribution function $f(q,\,p,\,t)$ (without subindex `$s$'). The number of particles in the box is $f(q,\,p,\,t)\,dqdp$. The change of the particle number is represented with the flux $f(q,\,p,\,t)\dot{q}$ and $f(q,\,p,\,t)\dot{p}$ like
\begin{eqnarray}
&&\frac{\partial\, f(q,\,p,\,t)}{\partial\, t}\,dqdp=\nonumber\\
&&f(q,\,p,\,t)\,\dot{q}\,dp-f(q+dq,\,p,\,t)\,\dot{q}\,dp+f(q,\,p,\,t)\,\dot{p}\,dq-f(q,\,p+dp,\,t)\,\dot{p}\,dq\nonumber\\
&&\sim-\dot{q}\frac{\partial f(q,\,p,\,t)}{\partial q}dpdq-\dot{p}\frac{\partial f(q,\,p,\,t)}{\partial p}dpdq.\label{Liouville1}\\
&&\Rightarrow \bigg(\frac{\partial f}{\partial t}+\dot{\bm q}\cdot \frac{\partial f}{\partial {\bm q}}+\dot{\bm p}\cdot \frac{\partial f}{\partial {\bm p}}\bigg)dpdq=0.
\label{Liouville2}
\end{eqnarray}
With more familiar symbols, Eq.(\ref{Liouville2}) is
\begin{eqnarray}
&&\frac{\partial f}{\partial t}+{\bm v}\cdot \nabla f+{\bm a}\cdot \nabla_V f=0\rightarrow
\frac{\partial f}{\partial t}+\nabla_{\Gamma}\cdot (f{\bm v})=0 \label{Liouville3} \\
&&\rightarrow \frac{\partial f}{\partial t}+\{f, \mathcal{H}\}=0. \label{Liouville4}
\end{eqnarray}
Here $\Gamma$ is a system state vector $(q,\,p)$, and $\{f, \mathcal{H}\}$ is Poisson bracket. And, they are related with $p=-\partial H/\partial q$ and $q=\partial H/\partial p$. This Liouville theorem indicates that the net change of density or distribution function in phase space is $zero$ as we move along the trajectory dominated by Hamiltonian dynamics. Mathematically, it is represented as the material derivative ($D f/Dt=\partial f/\partial t+\mathbf{v}\cdot\nabla f$) in phase space of $(q,\,p)$. The change of the system variable results in that of the distribution function modifying permittivity.\\}

In comparison to the overall distribution $f(\mathbf{r},\, \mathbf{v},\, t)$, the slightly higher (or lower) density electrons surrounding the nucleus can be regarded as the perturbed distribution {$f_1\,(=f-f_0)$}{\footnote{The perturbation of $f$ is not necessarily caused by the magnetic field alone. Due to the internal thermal energy in the system, the distribution of particles naturally perturbs. With the fixed magnetic field $B_0$, we get $f=f_0+f_1+f_2+...=...+q(v_0\times B_0+v_1\times B_0+v_2\times B_0)+...$.}}. Moreover, since the electrons shield the electric field from the nucleus, they effectively act as bound charges $\rho_b=\int f_1(\mathbf{r},\,\mathbf{v},\,t) d\mathbf{r}d\mathbf{v}$ and polarize the system with a dipole moment $\mathbf{P}$. Then, by utilizing the convolution property of Fourier Transformation and taking its divergence, we can separate the longitudinal permittivity $\epsilon_l$ from the electric displacement field $\mathbf{D}=\epsilon {\bf E}$ as follows \cite{1981phki.book.....L}:
\begin{eqnarray}
&&\nabla\cdot (\epsilon \mathbf{E})=\nabla\cdot (\epsilon_0 \mathbf{E}) + \nabla\cdot \mathbf{P}\\
&&\rightarrow k \epsilon_l(k,\,\omega)E(k,\,\omega)=k\,\epsilon_0E(k,\,\omega)-\rho_b(k,\,\omega)=k\,\epsilon_0E(k,\,\omega)+e\int f_1({k},\,\omega,\,\mathbf{v})\, d\mathbf{v}.
\label{D_field}
\end{eqnarray}
We apply this relation to the system that is weakly magnetized with the field strength $B_0$. The perturbed distribution function $f_1$ is \footnote{Harris dispersion relation is also obtained with this Vlasov equation and Poisson equation $\nabla^2 \Phi = -\rho/\epsilon_0 \rightarrow \epsilon_0\nabla^2 \Phi + q\int f_1 d{\bf v}=0 \rightarrow D(k,\,\omega)\Phi(k,\,\omega)=0$. Without polarization effect $P$ around the nucleus, dispersion relation between $k$ and $\omega$ is derived with the condition of nontrivial potential: $\Phi(k,\,\omega)\neq0\rightarrow D(k,\,\omega)\rightarrow \epsilon(k,\,\omega)=0$. Specifically, $k_{\|}=0$ in Harris dispersion relation is called Bernstein mode. In contrast, displacement field $\epsilon {\bf E} = \epsilon_0 {\bf E}+{\bf P}$ yields the nontrivial longitudinal permittivity $\epsilon(k,\,\omega)$. {In the consideration of current density, $\partial P / \partial t\rightarrow J_d\rho v=q\int vf_1\,dv$ is necessary instead of $\nabla \cdot {\bf P}=-q\int f_1\, dv$. This explains the reason why the permittivity in Eq.(\ref{BergmannEpsilon}) and Eq.(\ref{final_epsilon_numerical}) are not precisely consistent. However, to obtain the potential barrier from the Boltzmann approach, an intermediate electromagnetic concept such as $\nabla \cdot \mathbf{P}$ or $\mathbf{J}_D$ for the nontrivial permittivity is required.}}
\begin{eqnarray}
\frac{\partial f_{1}}{\partial t}+{\bf v}\cdot \nabla f_{1}-\frac{e}{m_e} {\bf E}\cdot \nabla _V f_{0}-\frac{e}{m_e} {\bf v}\times {\bf B}_0\cdot \nabla _V f_{1}=0.
\label{Boltzmann1}
\end{eqnarray}
We follow the standard plasma physics to solve this equation \citep{2017ipp..book.....G}. Using $v_x=v_{\perp}\cos\,\phi$, $v_y=v_{\perp}\sin\,\phi$, and cyclotron frequency $\omega_{ce}\equiv eB_0/m_e$, we can convert the fourth term, i.e., Lorentz force into $\omega_{ce}\partial f_{1}/\partial \phi$. Then, the Fourier transformed Boltzmann equation is represented as
\begin{eqnarray}
\frac{\partial f_{1}}{\partial  \phi}-i(\alpha+\beta \cos\,\phi)f_{1}+\frac{e}{m_e\omega_{ce}} {\bf E}\cdot \nabla _V  f_{0}=0,\label{Boltzmann3}
\end{eqnarray}
where $\alpha\equiv  (k_{\|} v_{\|}-\omega) / \omega_{ce}$ and $\beta\equiv  k_{\perp} v_{\perp} / \omega_{ce}$. And then, we get
\begin{eqnarray}
f_{1}=-\frac{e}{m_e\omega_{ce}}\,e^{i(\alpha\phi+\beta \sin\,\phi)}\int^{\phi} e^{-i(\alpha\phi'+\beta \sin\,\phi')}\,{\bf E}\cdot \nabla_Vf_{0}\,d\phi'.
\label{Boltzmann4}
\end{eqnarray}

Applying $f_1$ to Eq.(\ref{D_field}), we can derive the magnetized permittivity as follows:
\begin{eqnarray}
\epsilon_l=\epsilon_0+\frac{i\epsilon_0}{k^2\omega_{ce}}\omega^2_{pe}\int v_{\perp}dv_{\perp}dv_{\|}d\phi\,\overbrace{e^{i(\alpha\phi+\beta \sin\,\phi)}}^A\int^{\phi} \overbrace{e^{-i(\alpha\phi'+\beta \sin\,\phi')}\,{\bf k}\cdot \nabla_VF_{0}\,d\phi'.}^B
\label{permittivity1}
\end{eqnarray}
Here, plasma frequency $\omega_{pe}$ is defined as $\sqrt{n_{e0}e^2/\epsilon_0m_e}$, and the volume element in cylindrical coordinate is $d^3v=v_{\perp}dv_{\perp}dv_{\|}d\phi$. Plasma standard process shows how to solve the integral part of this equation (Eqs.(\ref{bimaxwellian})-(\ref{permittivity4}), refer to \cite{2017ipp..book.....G}). And, we use the anisotropic Maxwell distribution $F_{0}=f_{0}/n_{e0}$ for the analytic calculation:
\begin{eqnarray}
F_{0}=\bigg(\frac{1}{2\pi k_BT_{\perp}}\bigg)\bigg(\frac{1}{2\pi k_BT_{\|}}\bigg)^{1/2}e^{-\frac{m_s}{2k_BT_s}(v^2_{\|}+v^2_{\perp})},
\label{bimaxwellian}
\end{eqnarray}
where the temperature depends on the direction of the motion. The exponential term in `$A$' in Eq.(\ref{permittivity1}) can be represented by Bessel function {$ e^{i(\alpha+\beta \sin\,\phi)}=\sum_{m=-\infty}^{\infty}J_m(\beta)e^{i(\alpha+m)\phi}$}, and $k\cdot \nabla_V$ is written as $k_{\|}\partial /\partial\,V_{\|}+k_{\bot}\partial /\partial\,V_{\bot}cos\,\phi$ (see \cite{2003phpl.book.....B, 2005mmp..book.....A} for the detailed standard process). Combined `$A$' and `$B$' turn out to be
\begin{eqnarray}
\sum_{m,\,n}J_m(\beta)J_n(\beta)\bigg[ik_{\|}\frac{\partial F_{0}}{\partial v_{\|}}\frac{e^{i(m-n)\phi}}{\alpha+n}+i\frac{k_{\perp}}{2}\frac{\partial F_{0}}{\partial v_{\perp}}\bigg\{\frac{e^{i(m-n+1)\phi}}{\alpha+n-1}
+\frac{e^{i(m-n-1)\phi}}{\alpha+n+1}\bigg\}\bigg].
\label{Bessel13}
\end{eqnarray}
The index `$n$' is a dummy variable, and $\int^{2\pi}_0 e^{i(m-n)\phi}d\phi$ is defined as Dirac delta function $2\pi\delta_{m,\,n}$. Using Bessel recurrence relation  $J_{n+1}(\beta)+J_{n-1}(\beta) = (2\pi / \beta) J_n(\beta)$, we can derive
\begin{eqnarray}
\frac{\epsilon_l}{\epsilon_0}=1+\frac{2\pi\omega^2_{pe}}{k^2}\int_{-\infty}^{\infty}dv_{\|}\int^{\infty}_0v_{\perp}\,dv_{\perp}
\sum_n\bigg[\frac{m_sk_{\|}v_{\|}}{k_BT_e} + \frac{nm_e\omega_{ce}}{k_BT_e}\bigg]\frac{F_{0}J_n^2(\beta)}{k_{\|}v_{\|}-\omega+n\omega_{ce}},
\label{permittivity4}
\end{eqnarray}
where $T_e$ is the electron temperature. Before going further to find out $\epsilon_l$, we introduce other approaches yielding explain permittivity.\\

\subsection{Comparison with other methods}
\label{Comparison_with_other_methods}
There have been a couple of methods to derive permittivity, each producing somewhat different results, both formally and physically. Additionally, permittivities obtained through conventional approaches are mostly frequency-dependent. These methods will be briefly introduced here.
\subsubsection{Harmonic oscillator and bound charges}
An electrically neutral material (plasma) can be regarded as an equivalent dipole system, i.e., a harmonic oscillator. The motion of an electron bound to the nucleus or molecule is described as
\begin{eqnarray}
m\frac{d^2x}{d\,t^2}+m\,\gamma \frac{d\,x}{d\,t}+m\,\omega^2_0x=q\,E\,\cos\,\omega\,t.
\label{epsilon_harmonic_oscillator1}
\end{eqnarray}

With $P=(\epsilon-\epsilon_0)E$, permittivity can be derived as follows \citep{2017inel.book.....G}:
\begin{eqnarray}
\epsilon=\epsilon_0+\frac{nq^2_e}{m_e}\sum_j \frac{f_j}{\omega_{0j}^2-\omega^2-i\gamma_j\omega}.
\label{epsilon_harmonic_oscillator2}
\end{eqnarray}
`$n$' is the number of charged particles per unit volume, and $f_j$ is the number of electrons bound to the nucleus or molecule $j$. As the result implies, permittivity is not a fixed constant but a quantity that varies from $-\infty$ to $\infty$.\\

\subsubsection{Momentum equation and current density}
The bulk motion of the unbound electrons forms a current flow. Permittivity with current density $\mathbf{J}=\rho \mathbf{u}$ can be calculated with momentum equation:
\begin{eqnarray}
m_e\frac{d{\bf u}}{dt}=-e{\bf E}-m_e\nu_m{\bf u}.
\label{electron_force_equation1}
\end{eqnarray}
Here, `$\nu_m$' indicates the collision frequency yielding a frictional effect. This collision frequency corresponds to the damping coefficient $\gamma$. If the plasma system is driven by the harmonic electric field $\mathbf{E}(t)=\mathbf{\tilde{E}}\,e^{-i\omega t}$, we get ${\bf u}(t)\rightarrow \mathbf{\tilde{u}}\,e^{-i\omega t}$, where
\begin{eqnarray}
\mathbf{\tilde{u}}=\frac{e}{m_e}\frac{\mathbf{\tilde{E}}}{i\omega-\nu_m}.
\label{electron_force_equation2}
\end{eqnarray}
Then, using
\begin{eqnarray}
\mathbf{\tilde{J}}_{net}&=&-i\omega \epsilon_0\mathbf{\tilde{E}} -en_0\mathbf{\tilde{u}}\nonumber \\
&=&-i\omega \epsilon_0\bigg [1-\frac{\omega_{pe}^2}{\omega^2+i\omega \nu_m}\bigg]\mathbf{\tilde{E}}.
\label{electron_force_equation3}
\end{eqnarray}
we find the relative permittivity
\begin{eqnarray}
\epsilon/\epsilon_0\equiv 1-\frac{\omega_{pe}^2}{\omega^2+i\omega \nu_m}\sim 1-\frac{\omega_{pe}^2}{\omega^2}.
\label{effective_permittivity}
\end{eqnarray}
The result shows that as $\omega$ increases, $\epsilon$ approaches free space permittivity $\epsilon_0$. An intriguing observation is that when $\omega$ is smaller than the plasma frequency $\omega_{pe}$ ($\omega < \omega_{pe}$), $\epsilon$ is negative. As $\omega$ approaches $zero$, $\epsilon$ becomes negatively divergent ($-\infty$). In contrast, $\epsilon$ converges to $\epsilon_0$ as $\omega\rightarrow \infty$.\\

{\subsubsection{Momentum equation and continuity equation (fluidic approach)}
On the other hand, Hatami \citep{2021NatSR..11.9531H} derived the sheath properties in active magnetized multi-component plasmas. He solved the static ($\partial /\partial t=0$) continuity equation and Navier-Stokes equation in a compressible plasma system with a fixed magnetic field. The equations were solved with some physically simplified assumptions, which is different from conventional fluid approaches such as Mean Field Theory or Eddy-Damped Quasi-Normalized Markovian Approximation \cite{1975PhT....28f..59L, 1990cp...book.....M, 2022ApJ..park}. Nonetheless, a consistent result of decreasing $|\phi|$ with the $B$ field is illustrated. Additionally, Salimullah et al. \cite{2008PhLA..372.2291S} derived $\epsilon\sim \mathcal{O}(1/B)$ using momentum and continuity equations.\\}

Permittivity $\epsilon$ and magnetic permeability $\mu$ decide the wave in the material. Various waves propagate in material with $\epsilon>0$ and $\mu>0$. But, for $\epsilon\mu<0$ there is no propagating wave, only the evanescent wave exists in principle\footnote{Exceptionally, $\epsilon < 0$ and $\mu > 0$ is known to propagate in the interface of plasma-vacuum or plasma-dielectric material.}. However, if $\epsilon<0$ and $\mu<0$ ($\epsilon\mu>0$), negative index material, waves can propagate in the material. Negative refractive index material is also called left handed or metamaterial because of the opposite direction of triple set of Faraday and Ampere's law \citep{1968SvPhU..10..509V}.
\subsubsection{Boltzmann equation and current density}
Bergman \cite{2000PhPl....7.3476B} applied kinetic approach (Vlasov equation) to Maxwell equation. He assumed that the spatially inhomogeneous electric field would yield the magnetic field through Faraday's law $\partial {\bf B}/\partial\,t=\nabla \times {\bf E}$, which in turn induces the current density and electric displacement field through Ampere's law $\nabla \times {\bf B}=\mu_0({\bf J}+\epsilon_0\partial {\bf E}/\partial t)$. And then, using Ohm's law, the author analytically derived permittivity tensor $\stackrel{\leftrightarrow}{\epsilon}$$(k,\,\omega,\,B)$ including the transverse $\epsilon_{\bot}$ and longitudinal $\epsilon_{\|}$. {The result is formalistic and needs additional physical assumptions for a practical use. For example, in a cold plasma ($v_{th, e}\rightarrow 0$), the longitudinal component $\epsilon_{33}$ and transverse components $\epsilon_{11}, \epsilon_{22}$ are represented as
\begin{eqnarray}
\epsilon_{11}/\epsilon_{0}&=&\epsilon_{22}/\epsilon_{0}=1-\frac{\omega^2_{pe}}{\omega^2-\omega^2_{ce}}\\
\epsilon_{33}/\epsilon_{0}&=& 1-\frac{\omega^2_{pe}}{\omega^2_{ce}}
\label{BergmannEpsilon}
\end{eqnarray}
The transverse permittivity depends on $\omega_{pe}$ and $\omega^2-\omega^2_{ce}$, so the effects of $n_{e0}$ and $B$ vary relative to $\omega$. However, the longitudinal permittivity is inversely proportional to $n_{e0}$ and directly proportional to $B$. In contrast to Eq.(\ref{Boltzmann1}) for the external ${\bf B}_0$ field, this self-consistent system utilizes internally generated magnetic fields. Consequently, the field scale is much smaller compared to $B_0$ and exhibits more interaction with the system.}

\subsection{Numerical calculation of longitudinal permittivity}
For the numerical calculation of Eq.(\ref{permittivity4}), we expand Bessel function to make the equation more suitable \citep{2005mmp..book.....A}.
\begin{eqnarray}
\frac{\epsilon_l}{\epsilon_0}&=&1+\frac{2\pi\omega^2_{pe}}{k^2}\int_{-\infty}^{\infty}dv_{\|}\int^{\infty}_0v_{\perp}\,dv_{\perp}
\sum_n\bigg[\frac{m_ek_{\|}v_{\|}}{k_BT_e} + \frac{nm_e\omega_{ce}}{k_BT_e}\bigg] \frac{F_{0}}{k_{\|}v_{\|}-\omega+n\omega_{ce}}\bigg(\sum_{s=0}^{\infty}\frac{(-1)^s}{s!(s+n)!}
\bigg(\frac{\beta}{2} \bigg)^{n+2s}\bigg)^2.
\label{permittivity7}
\end{eqnarray}
Technically, longitudinal permittivity $\epsilon_l$ in this equation represents the area between the horizontal axis of $v_{\|}$ and integrand. However, the typical residue theorem with singularities cannot be applied because of the divergent $F_{0}$ with $v_{\|, im}\rightarrow \pm i\infty$ (see $F_0\sim exp[-\frac{m_e}{2k_BTe}(v^2_{\|}+v^2_{\bot})]$). Instead, we should integrate its principal value and poles directly.
\begin{eqnarray}
\frac{\epsilon_l}{\epsilon_0}&=& 1+\frac{2\pi\omega^2_{pe}}{k^2}P\int_{-\infty}^{\infty}dv_{\|}
\int^{\infty}_0v_{\perp}\,dv_{\perp}\nonumber\\
&&\bigg(\sum_n \frac{m_e}{k_BT_e}\bigg(\frac{k\, v_{\|}\cos\theta
+ n\omega_{ce}}{k\, v_{\|}\cos\theta-\omega+n\omega_{ce}}\bigg)F_{0}\bigg[\sum_{s=0}^{\infty}\frac{(-1)^s}{s!(s+n)!}
\bigg(\frac{m_ekv_{\perp}\sin\theta}{2eB_0}\bigg)^{n+2s}\bigg]^2\bigg)\nonumber\\
&& + i\pi \frac{k}{|k|}\frac{2\pi\omega^2_{pe}}{k^2}
\sum_n\frac{m_e}{k_BT_e}\big(k\, v_{\|}\cos\theta+nm_e\omega_{ce}\big)\bigg[\sum_{s=0}^{\infty}\frac{(-1)^s}{s!(s+n)!}
\bigg(\frac{m_ekv_{\perp}\sin\theta}{2eB_0} \bigg)^{n+2s}\bigg]^2F_{0}.
\label{final_epsilon_numerical}
\end{eqnarray}
We applied the trapezoidal rule to calculate Eq.(\ref{final_epsilon_numerical}) numerically \citep{2012phpl.book.....M}. The wavenumber $k$ ranges from $1$ to $3000$, $v$ ranges from $v_{min}=-10^8$ to $v_{max}=10^8$, and the mesh size is $\Delta v=0.5$. {In principle, $k\rightarrow \infty$, $|v|\rightarrow \infty$, and $\Delta v$ should be $\sim 0$. However, these environments mixed with the huge ($\pm c$) and tiny figures ($e,\,m_e,\, k_B$, etc.) are numerically unfriendly. Therefore, we chose $\Delta v$ and $|v|$ to ensure that $\int^v_{-v} F_{0} dv=1$. Other integrands were attached to $F_{0}$ and calculated together. If $|v|$ is scaled to be a unit, $\Delta v$ is in the order of $10^{-9}$. In this case, the re-scaled integrands including $F_{0}$ yield additional numerical errors with a unreliable result.} A more advanced numerical scheme does not seem to be necessarily required{\footnote{{In principle, as $|v|\rightarrow \infty$ and $\Delta v\rightarrow 0$, numerical calculations are likely to fail. However, to address this issue and ensure computational reliability, we adopted an optimized condition. Specifically, we set $|v|$ to one-third of the speed of light and $\Delta v$ to 0.5. Under these conditions, the simulation yielded a normalized Maxwell integration: $\int^v_{-v}F,dv=1$. Additionally, we calculated the Maxwell distribution function considering other asymmetric and complex terms.}}}. And, we determined the electron density $n_{e0}=8.18363\times 10^{13}m^{-3}$ and the temperature $T_e=2.38\times 10^6\,K$ near the Solar tachocline regime(0.7$R_\odot$), with an arbitrary frequency $\omega=10^4 Hz$ smaller than the plasma frequency $\omega_{pe}=5.1\times 10^8 Hz$. We used openmp with 64 cores (128 threads) for the parallel computation.\\

The integrand has singular points $v_{res,\,n}$ that makes the denominator $zero$. Landau\cite{1965445} suggested that energy transport between particles and waves should take place at the singularity (Landau damping). However, we do not consider their interaction that is beyond the scope of our work. Therefore, we divided the integral range of $v_{\|}$ into two intervals: ($v_{min},\,v_{res,\,n}-\delta$) and ($v_{res,\,n}+\delta,\,v_{max}$), excluding the singular point $v_{res,\,n}$. {Also, theoretically, $\delta$ should approach $zero$, but it is not practically well defined. Near the singularity, the integration does not converge before encountering computational roundoff errors. Therefore, we chose an arbitrarily small value of $\delta = 500 nm$, which corresponds to the boundary wavelength between visible light and infrared. {Landau suggested energy transport between plasma particles and the wave at $v=\omega/k$, implying $\delta\sim 0$. However, this value is numerically impractical. There is no convergence as $\delta$ decreases to values near zero, within the limits of roundoff error.  } A real experiment is necessary to determine the distance at which energy transport begins and the amount of energy exchanges. } Finally, we expanded Bessel function up to $v^{18}_{\perp}$ for the case that the wavenumber is almost parallel to the $B$ field, i.e., $\beta\sim sin\,\theta\sim0$. But, the result was already saturated in the order of $v^{10}_{\perp}$$:\epsilon(v^{10}_{\perp})\sim \epsilon(v^{18}_{\perp})$.\\

\section{Numerical result}
Fig.~\ref{f1} illustrates the Fourier-transformed evolving permittivity $\epsilon$ ($=\epsilon_0\epsilon_r$) under the influence of the magnetic field. We calculated Eq.(\ref{final_epsilon_numerical}) with discrete values of $k$ ($1-3000$) and magnetic field strength ($B=0-1\times 10^{-5} \text{G}$). The permittivity remains degenerate up to a critical wavenumber $k_{\text{crit}}$, which depends on the strength of the magnetic field. Beyond this critical point, it begins to deviate and separate as $k > k_{\text{crit}}$. The amplification of $\epsilon$ is inversely proportional to the magnetic field strength. For magnetic fields stronger than $1\times 10^{-7}\, \text{G}$ as indicated by the dashed line, the permittivity is practically the same as in the non-magnetized case. This is technically attributed to the presence of the $B$ term in the denominator of Eq.(\ref{final_epsilon_numerical}). In physical terms, a weaker magnetic field induces a larger cyclotron radius ($r\sim 1/B$), increasing the superposition of electrons, which results in the enhanced $f_1$. Additionally, the results suggest that the magnetic field interacts more efficiently with shorter wavelengths i.e., larger values of $k$.\\

Fig.\ref{f21} presents the evolution of $\epsilon(k)$ as a function of temperature with an applied magnetic field of $4.5\times 10^{-9}\mathrm{G}$. The plot demonstrates that permittivity is inversely proportional to temperature, similar to its relationship with the magnetic field $\epsilon\sim 1/T$. However, their underlying mechanisms are fundamentally different. Higher temperatures lead to greater dispersion in the electron distribution function, approximately following $\exp(-m_ev^2/k_BT)$, resulting in a flatter distribution. This slowly changing the flat distribution leads to the increased symmetry in the integrand. Consequently, the area between the integrand and the horizontal axis decreases, resulting in a reduction in permittivity. \\ 

Fig.\ref{f22} illustrates the impact of electron density $n_{e0}$ on permittivity. In comparison to the effects of magnetic field or temperature, $n_{e0}$ tends to increase permittivity proportionally $\epsilon\sim n_{e0}$. This is in contrast to Eq.(\ref{effective_permittivity}) and Eq.(\ref{BergmannEpsilon}), which are derived under the assumption of unbound current flow.\\ 

Fig.~\ref{f2} presents the inverse Fourier-transformed $\epsilon_r(r/\lambda_D)$ derived from $\epsilon(k)$ in Fig. \ref{f1}. Here, `$r$' represents the distance normalized with the Debye length $\lambda_D (=\sqrt{\epsilon_0k_BT/e^2n_e}\approx 1.17\times 10^{-2}\text{cm})$. The nucleus is located at $r=0$, and its charge is represented by $Q\delta(r)$. To obtain $\epsilon_r(r)$, we performed the inverse Fourier transformation of $\epsilon(k)$ from Fig. \ref{f1} using the following formula:
\begin{eqnarray}
\epsilon(r_n)=\frac{1}{N}\sum_{k=0}^{N-1}\epsilon(k)\exp\left(i\frac{2\pi k n}{N}\right),
\label{IDFT1}
\end{eqnarray}
where $N=3000$ and $n/N=r_n/\lambda_D$. {Permittivity is numerically calculated through the discrete Fourier transformation of Eq.(\ref{Boltzmann1}), (\ref{permittivity7}), (\ref{final_epsilon_numerical}) using the discrete wavenumbers. Essentially, Fourier transformation (FT) assumes periodic values or functions. However, since FT involves integrating arbitrary functions to transform them into another domain, it can effectively decompose the internal structure of a function into wavenumber or frequency components, irrespective of whether the function exhibits periodicity or not (Mathematical methods for physicists by Arfken, chapter 15.2, 5$^{th}$ edition).} Close to the nucleus, $\epsilon$ exhibits distinct levels corresponding to the applied magnetic field. Permittivity is inversely proportional to the magnetic field. Similar to $\epsilon(k)$, above a critical field $B_{\text{crit}}$, permittivity is no longer split but converges to that of the non-magnetized system. At $n\sim N-1$, $\cos (2\pi k n/N)$ is almost equal to 1, which results in a sudden increase in $\epsilon$ at $r \sim \lambda_D$. {The permittivity here is attributed to the electric field emanating from the nucleus, whereas the permittivity in Eq.~(\ref{BergmannEpsilon}) arises from the induced electric field generated by the perturbed magnetic field. This represents more of an additional adjustment to the permittivity rather than its fundamental form.}\\

Fig. \ref{f3} illustrates the evolving potential energy, denoted as $\phi=Q/4\pi \epsilon r$, for a hydrogen nucleus with permittivity $\epsilon(r)$. Since permittivity plays a crucial role in the denominator, potential energy changes in proportion to the magnetic field and approaches the non-magnetized potential as it surpasses the critical magnetic field. The penetration factor $P(E)$ is determined based on this potential energy.\\

Fig. \ref{f4} depicts how $P(E)$ changes with the magnetic field. The weak magnetic field decreases the potential barrier, increasing the likelihood of penetration and consequently boosting the reaction rate. In principle, the actual potential barrier should encompass the interaction energy among the screening charges surrounding the two interacting nuclei and the surrounding lighter nuclei. However, we do not delve into these complex effects as they are beyond the scope of this paper. In Fig.~1 and 2, we demonstrated the permittivity and potential energy derived using the Boltzmann method. However, it would be more appropriate to compare this statistical approach with a different method.\\

{On the other hand, in section. (\ref{Comparison_with_other_methods}), we have introduced a few other methods to determine permittivity (also see Table 1). It is worthwhile to consider whether those approaches, particularly Bergman's method \cite{2000PhPl....7.3476B}, can be applied to the potential barrier for nucleosynthesis. However, to put it briefly, it is not suitable for nuclear reactions. The potential barrier originates from the electric field $E$ of the nucleus charge. However, Bergman's approach assumes that the $E$ field is induced by the magnetic field $\partial {\bf B}/\partial t = -\nabla \times {\bf E}$. An electric field like this is not directly related to the potential barrier. This simple criterion can be applied to other approaches.\\}

{Figs.\ref{f41}, \ref{f42} compare unmagnetized Debye potential barriers $\Phi_D$(solid line)
\begin{eqnarray}
\Phi_D=\frac{Q}{4\pi \epsilon_0r}\exp\bigg[-\frac{\sqrt{2}r}{\lambda_D}\bigg],\,\, \epsilon\rightarrow \epsilon_0\exp\bigg[r \sqrt{\frac{2n_e e^2}{\epsilon_0k_BT_e}}\bigg],
\label{debye_potential1}
\end{eqnarray}
and unmagnetized Boltzmann potential barrier $\Phi_{B}$(other lines) from
\begin{eqnarray}
f_{1}=\frac{e}{i\,m_e}\frac{{\bf E}\cdot \nabla_Vf_{0}}{({\bf k}\cdot{\bf v}-\omega)}.
\label{Boltzmann_no_B}
\end{eqnarray}
Despite the same $T=2.38\times 10^6 K$ and $n_{e0}=8.18\times 10^{13}m^{-3}$, there is a quantitative discrepancy between these two potentials. However, as $T$ increases or $n_{e0}$ decreases, $\Phi_{B}$ grows to approach $\Phi_D$. Similarly, Fig.~\ref{f5} and \ref{f6} depict the dependence of $\Phi_D$ on temperature and electron density. As $T$ decreases or $n_{e0}$ decreases, $\Phi_{D}$ decreases to approach $\Phi_B$.\\

A closer examination of Fig. \ref{f41}, \ref{f5} reveal that $\Phi_B$ and $\Phi_D$ exhibit a proportional dependence on temperature. In contrast, Fig. \ref{f42}, \ref{f6} show an inverse proportionality between $\Phi_B$ \& $\Phi_D$ and $n_{e0}$. To achieve similarity between $\Phi_B$ and $\Phi_D$, it is necessary to control either $T$ or  $n_{e0}$. This discrepancy arises from the different approaches used in the statistical distribution function and the averaged momentum (Langevin) equation. Nonetheless, the qualitatively consistent dependence on $n_{e0}$ or $T$ in both methods implies the validity of statistical Boltzmann approaches.}\\

{In Fig. \ref{f7} and \ref{f8}, we compared the magnetized potential $\Phi_{B,B}$ (solid line, Eq. (\ref{final_epsilon_numerical})) and magnetized Debye potential $\Phi_{D,\,B}$ (other lines, Eq.(\ref{derivation_of_new_Debye_potential_12})). Additionally, the unmagnetized Debye potential $\Phi_{D}$ is overlapped (diamond). Fig.\ref{f7} illustrates that $\Phi_{D,B}$, as well as $\Phi_{B,B}$, evolves proportionally with the magnetic field. And, Fig.\ref{f8} shows that the potential barrier is also proportional to the temperature like the conventional Debye potential $\Phi_D$ and Boltzmann method $\Phi_{B}$. However, $\Phi_{D,B}$ and $\Phi_{B,B}$ are inversely proportional to $n_{e0}$, which is physically reasonable. And, the unmagnetized $\Phi_{D}$ coincides with $\Phi_{D, B}$ for $B=4\times 10^{-9}G$. This indicates that $\Phi_{D, B}$ with this $B$ field is already saturated (see Eq.(\ref{derivation_of_new_Debye_potential_12})). On the other hand, near the nucleus ($r\sim0$), $\Phi_{B,B}$ exhibits a pronounced ridge-like feature in the vicinity of the nucleus instead of the monotonic evolution. This is associated with the effect of short wavelength (large $k$) interacting with the magnetic effect. Besides, the consistent influence of $n_{e0}$ on $\Phi_{D,B}$ also exists, but we did not include this effect.}\\

\section{Theoretical Analysis II: Magnetized Debye potential}
We now consider the magnetic effect on the conventional Debye potential. {We assume an isotropic and homogeneous plasma system, which is valid with a weak magnetic field. Here, there is no exact critical $B$ field. But as its strength increases, the anisotropy in the system also grows.} And, we add Lorentz force and the (electromagnetic) collision frequency $\nu_m$ to Langevin (momentum) equation.

For the nucleus, the momentum equation is
\begin{eqnarray}
&&m_in_i\frac{d\mathbf{U}_i}{dt} = n_ie({\bf E}+{\bf U}_i\times {\bf B})-k_BT\nabla n_i-\nu_m m_in_i{\bf U}_i.\\
&&\Rightarrow -\frac{e}{m_i}\nabla \phi + \frac{e}{m_i}U_iB-\frac{k_BT}{m_i}\frac{\nabla n_i}{n_i}-\nu_mU_i + i\omega {U}_i\sim 0.
\label{derivation_of_new_Debye_potential_1}
\end{eqnarray}
And, the equation for the electron is
\begin{eqnarray}
\frac{e}{m_e}\nabla \phi - \frac{e}{m_e}U_eB-\frac{k_BT}{m_e}\frac{\nabla n_e}{n_e}-\nu_mU_e + i\omega {U}_e\sim 0.
\label{derivation_of_new_Debye_potential_2}
\end{eqnarray}
To derive $n_i(r)$ and $n_e(r)$, we integrate these equations over distance from infinity to $r$ assuming quasi-continuous velocity distribution (mean value theorem):
\begin{eqnarray}
&&n_e(\infty)=n_i(\infty)\equiv n_0,\, \int^r_{\infty}\nabla \phi\,dr=\Phi(r),\,  \int^r_{\infty}U_idr \rightarrow \overline{U}_i\overline{r}_i,\,\,\int^r_{\infty}U_edr \rightarrow -\overline{U}_e\overline{r}_e.
\label{derivation_of_new_Debye_potential_3}
\end{eqnarray}
For the ion, we have
\begin{eqnarray}
\ln\frac{n_i}{n_0}=-\frac{e}{k_BT}\Phi+(eB-m_i\nu_m+im_i\omega)\frac{\overline{U}_i\overline{r}_i}{k_BT}.
\label{derivation_of_new_Debye_potential_4}
\end{eqnarray}
With the application of cyclotron motion, the Lorentz force is valanced with the centrifugal force resulting in $\overline{r}_i\sim m_i\overline{U}_i/eB$. Also, we can apply the thermal energy relation $3n_0m_i\overline{U}^2_i=k_BT$. Then, ion density is
\begin{eqnarray}
\ln\frac{n_i}{n_0}&=&-\frac{e\Phi}{k_BT}+(eB-m_i\nu_m+im_i\omega)\frac{1}{3n_0eB}.\\
\rightarrow n_i&=&n_0\,exp\bigg[-\frac{e\Phi}{k_BT} + \frac{(eB-m_i\nu_m+im_i\omega)}{3n_0eB}\bigg].
\label{derivation_of_new_Debye_potential_5}
\end{eqnarray}
Next, we derive $n_e$ in the same way:
\begin{eqnarray}
n_e&=&n_0\,exp\bigg[\frac{e\Phi}{k_BT} + \frac{(eB+m_e\nu_m-im_e\omega)}{3n_0eB}\bigg].
\label{derivation_of_new_Debye_potential_6}
\end{eqnarray}
Using Poisson eqution $-\epsilon_0 \nabla^2 \Phi=e(n_i-n_e)+Q\delta (r)$, we get $n_i-n_e$ is as follows:
\begin{eqnarray}
n_i-n_e\sim n_0\bigg( -\frac{2e\Phi}{k_BT}-\frac{(m_i+m_e)(\nu_m-i\omega)}{3n_0eB}\bigg),
\label{derivation_of_new_Debye_potential_7}
\end{eqnarray}
where $m_e$ can be neglected. The differential equation to be solved is
\begin{eqnarray}
\frac{1}{r^2}\frac{\partial}{\partial \,r}\bigg(r^2 \frac{\partial\,\Phi}{\partial\,r}\bigg)-\frac{2}{\lambda^2_D}\Phi-\frac{m_i}{3\epsilon_0B}\,(\nu_m-i\omega)=0.
\label{derivation_of_new_Debye_potential_8}
\end{eqnarray}
With a trial function $\Phi=eF(r)/4\pi \epsilon_0 r$, we have
\begin{eqnarray}
\frac{\partial^2F(r)}{\partial r^2}-\frac{2}{\lambda_D^2}F(r)=\frac{4\pi(\nu_m-i\omega)m_ir}{3eB}.
\label{derivation_of_new_Debye_potential_9}
\end{eqnarray}
{This second order inhomogeneous differential equation of $F(r)$ can be analytically solved:}
\begin{eqnarray}
F(r)=c_1e^{\frac{-\sqrt{2}r}{\lambda_D}} + c_2e^{\frac{\sqrt{2}r}{\lambda_D}} + \frac{2\pi r m_i\lambda^2_D(i \omega - \nu_m)}{3eB}
\label{derivation_of_new_Debye_potential_9a}
\end{eqnarray}
Now we can derive Potential function:
\begin{eqnarray}
\Phi(r)=c_1\frac{Q}{4\pi \epsilon_0 r}e^{-\frac{\sqrt{2}r}{\lambda_D}}+c_2\frac{Q}{4\pi \epsilon_0 r}e^{\frac{\sqrt{2}r}{\lambda_D}}+\frac{m_i\lambda^2_D}{6\epsilon_0 B}(i\omega-\nu_m).
\label{derivation_of_new_Debye_potential_10}
\end{eqnarray}
The result demonstrates that the potential energy comprises two components: one arising from the nucleus and the other from the magnetic field and collision effect, denoted as $\nu_m$. $\nu_m$ is influenced by the combined effects of the electric field, magnetic field, and thermal pressure. Consequently, the collision frequency may be confined to the inner range of $\lambda_D$. Therefore, to ensure that $\Phi(\infty)=0$, the value of $c_2$ should be
\begin{eqnarray}
c_2=-\frac{4\pi r}{Q}e^{-\sqrt{2}r/\lambda_D}\frac{i\omega\,m_i\lambda^2_D}{6B}.
\label{derivation_of_new_Debye_potential_11}
\end{eqnarray}
If we set $c_1=1$, the modified Debye potential is
\begin{eqnarray}
\Phi_{D,B}=\frac{Q}{4\pi \epsilon_0 r}e^{-\frac{\sqrt{2}r}{\lambda_D}}-\frac{\nu_mm_ik_BT}{6n_0e^2B}.
\label{derivation_of_new_Debye_potential_12}
\end{eqnarray}
This result demonstrates that the potential is directly proportional to the magnetic field $B$, which is consistent with the kinetic model (Fig. \ref{f3}). $\Phi_{D,B}$ also suggests that a stronger magnetic field beyond a critical strength $B_{\text{crit}}$ paradoxically negates its effect on the potential energy. {Technically, the potential barrier is inversely related to the collision frequency $\nu_m$, which establishes a connection between the magnetic field and the centrifugal force in Debye potential. Besides, the collisional effect makes the system isotropic and homogeneous by transporting momentum in all directions. And it prevents the potential energy from diverging at infinity.}  Currently, there is no exact $\nu_m$ known for Debye shielding. Therefore, we have referred to the electron-charge collision frequency derived from Coulomb collision cross-section \citep{2017ipp..book.....G}:
\begin{eqnarray}
\nu_{m}=\frac{n_0e^4\ln\Lambda}{32(\pi m_e)^{1/2} \epsilon_0^2 (2k_BT)^{3/2}}.
\label{collision_frequency_0}
\end{eqnarray}
Plasma collision parameter $\Lambda$ is defined as $12\pi n_0\lambda_D^3$, and the Coulomb logarithm $\ln \Lambda$ typically falls in the range of 10 to 40 for most plasma systems. In our case, it is approximately $22.3$. It's important to note that the effects of $T$ or $n_0$ on the Debye length $\lambda_D$ are implicitly considered. For example, Eq. (\ref{derivation_of_new_Debye_potential_12}) indicates that $\Phi$ appears to be inversely proportional to $T$, but Fig. \ref{f8} demonstrates that increasing temperature $T$ actually raises $\Phi$. This observation is consistent with $\Phi_{B,B}$ as well.\\

\section{Summary}
In our study, we tackled the problem of weak magnetization by solving the Boltzmann equation within the framework of an isolated canonical ensemble, which consists of the nucleus and bound charges. Our investigation revealed an intriguing relationship: permittivity exhibited an inverse proportionality to the strength of the magnetic field. This finding indicated that the potential barrier governing the fusion of two nuclei evolved in tandem with the magnetic field's intensity. Such a result can be associated with Liouville's theorem, a fundamental concept in physics, which states that the net change in density or distribution function in phase space is zero when following a trajectory governed by Hamiltonian dynamics.\\

The presence of a weak magnetic field had a distinct impact on the Boltzmann equation. It effectively reduced the acceleration effect in the equation, which, in turn, led to a decreased constraint on electrons imposed by the magnetic field. Consequently, there was an increase in the fluctuating electron distribution, denoted as $f_1(\mathbf{r},\,\mathbf{v},\,t)$ in configuration space, to compensate for the loss. The equation $\nabla \cdot (\epsilon {\bf E})=\epsilon_0 \nabla \cdot {\bf E}+e\int f_1d\mathbf{r}d\mathbf{p}$ elucidates how the growth of $f_1$ contributes to the rising permittivity, subsequently causing a decrease in the potential barrier. However, it's worth noting that for magnetic fields surpassing a critical threshold, the electrons become strongly constrained, resulting in behavior akin to a non-magnetized system.\\

On the other hand, the statistical Boltzmann approach comes with its limitations, primarily due to incomplete Fourier transformations and constraints inherent to the applied statistical theory. To assess the validity of the Boltzmann method, we conducted a comparison between the unmagnetized Boltzmann model and the conventional Debye potential without a magnetic field. While there exists a quantitative discrepancy between the two, their qualitative consistency lends support to our analysis. Building upon this, we derived the magnetized Debye potential energy $\Phi_{D,B}$ as a function of temperature, charge density, and magnetic field. Notably, the influence of the magnetic field on the system was not direct but rather indirect, coming into play when thermal kinetic energy, centrifugal force, and Lorentz force found a delicate balance.\\

We proceeded to compare the potential energy obtained from the Boltzmann potential $\Phi_{B,B}$ with $\Phi_{D,B}$, and, as expected, there was a quantitative discrepancy. However, both approaches exhibited consistent trends in relation to the magnetic field, temperature, and the electron density. In both cases, a weak magnetic field played a significant role in reducing the potential barrier, leading to a substantial increase in the penetration factor and reaction rate.\\

Finally, the figures and calculations in this paper are related to the potential alterations in permittivity and potential barriers within the solar core of which physical conditions can be inferred. However, these plots and calculations, generated using data from the early Universe, encompassing temperature, density, and magnetic fields, can also illustrate changes in permittivity and potential barriers. Such changes would result in enhanced nuclear reaction rates, offering deeper insights into nucleosynthesis influenced by the pervasive magnetic fields in the Universe since the epoch of hot Big-Bang expansion until today throughout the Galactic and stellar evolution. Beyond nucleosynthesis, the ability to maintain the plasma's neutral state and control the Coulomb potential through magnetic fields can have broader applications, including chemical bonding and electrical conductivity.\\

\begin{acknowledgements}
K. Park acknowledges the support from National Research Foundation of Korea:NRF-2021R1I1A1A01057517, NRF-2020R1A2C3006177, NRF-2021R1A6A1A03043957, and NRF-2020R1F1A1072570. Y. Luo is supported by the Boya Fellowship of Peking University and the National Natural Science Foundation of China (12335009). T. Kajino was supported in part by the National Natural Science Foundation of China (No. 12335009), the National Key R\&D Program of China (2022YFA1602401) and Grants-in-Aid for Scientific Research of Japan Society for the Promotion of Science (20K03958).
\end{acknowledgements}

\bibliography{bibfile_2024_0213}

\end{document}